\documentstyle[psbox,PASJadd]{PASJ95}
\newcommand{\vol}{}
\newcommand{\no}{}
\newcommand{\lett}{}
\newcommand{\spage}{}
\newcommand{\rdate}{1999 December 27}
\newcommand{\adate}{2000 April 20}

\begin{document}

\title{Discovery of a New Supernova Remnant in the Direction of G69.7+1.0}
\author{Kumi {\sc Yoshita}, Emi {\sc Miyata}, and Hiroshi {\sc Tsunemi}\\
{\it Department of Earth and Space Science, Graduate School of
Science, Osaka University} \\
{\it 1-1 Machikaneyama, Toyonaka, Osaka 560-0043}\\
{\it E-mail(KY): kyoshita@ess.sci.osaka-u.ac.jp}}

\abst{
We discovered a middle-aged supernova remnant 
in the vicinity of G69.7+1.0 using the ASCA satellite.
G69.7+1.0 was identified in the 2.7~GHz survey and 
classified as a shell-type SNR
with a diameter of 16$^{\prime}$.
During the ROSAT all-sky survey,
the X-ray emission was detected in the direction of G69.7+1.0.
However, it extends beyond the radio shell, 
and an X-ray bright region was located outside of the radio shell.
A spectral study with the ASCA and ROSAT shows 
a thin thermal plasma with an electron temperature of $\sim$ 0.4~keV.
There is no significant variation of the spectral parameters 
over the field of view, except for the lower column density of the
eastern part.
We also found a large shell structure which surrounds
the X-ray bright region in both optical and radio images.
We suggest that the observed X-ray emission is associated with
the large optical and radio shell, and
that they are part of a new SNR, different from the radio SNR G69.7+1.0,
which we have named AX~J2001$+$3235 or G69.4$+$1.2.
The large shell and the electron temperature of $\sim$ 0.4~keV 
indicate that AX~J2001$+$3235 is an evolved SNR.
From a comparison with the column density of CTB 80 (G69.0+2.7), 
we estimate that the distance of the SNR is about 2.5~kpc.
}

\kword{ISM: individual (G69.7+1.0) --- supernova remnants --- X-rays: ISM}

\maketitle
\thispagestyle{headings}

\section{Introduction} 
Most galactic supernova remnants (SNRs) have been 
discovered and classified by observations in the radio band.  
Green's present catalog (1998 September version; Green 1998) 
lists 220 SNRs.
These remnants are classified into three types:
``shell-type'', ``filled-center'', and ``composite''.
Except for filled-center SNRs, they have radio shells.
While some SNRs classified as shell-type in radio also have shells in X-rays, 
others show a centrally-brightened X-ray morphology.

SNRs are also classified according to their age.
While Cassiopeia A and Tycho's SNR,
which are $\sim$ 300 yr (Fabian et al.\ 1980) and 
430 yr old (Reid et al.\ 1982),
are classified as young shell-type remnants,
the Cygnus Loop, Vela, and Puppis SNRs with ages of between 4000 and 20000 yr
(Ku et al.\ 1984; Gorenstein et al.\ 1974),
are classified as evolved, middle-aged remnants.
These evolved remnants typically appear as large radio shells
containing many optical filaments.
The radio emission is synchrotron radiation, which occurs 
when relativistic electrons move through a magnetic field, and 
the radio shell is considered to trace the blast wave.
The X-ray emission comes from a thin thermal plasma heated either by the
blast wave or by the reverse shock. 
Therefore, we can recognize evolved SNRs as
X-ray emission well within a large radio shell 
(Seward et al.\ 1990; Green 1998).

The SNR G69.7+1.0 was discovered in the 2.7 GHz survey.
The radio spectrum has an energy index, $\alpha$ 
(${\rm S}_{\nu}\sim\nu^{-\alpha}$ ), of 0.8 (Reich et al.\ 1988)
and it is classified as a shell-type SNR with a diameter of 16$^{\prime}$.
A $\Sigma$--{\it D} relation suggests the distance to G69.7+1.0 to be 14.4~kpc 
(Case, Bhattacharya 1998).

In X-rays, the ROSAT all-sky survey discovered a bright X-ray arc in 
the direction of G69.7+1.0 (Asaoka et al.\ 1996).
It was observed in 1993 October with the ROSAT 
position-sensitive proportional counter (PSPC) in the pointing mode.
The spectral study of this X-ray arc has not been made yet, 
which can clarify its nature.
In this paper, we present the results 
obtained with the ASCA satellite combined with the ROSAT PSPC data.

\section{Observations and Data Analysis}

% Table 1
\begin{table*}[t]
\begin{center}
Table~1.\hspace{4pt}Summary of ASCA observations.\\
\end{center}
\vspace{6pt}
 \begin{tabular*}{\textwidth}{@{\hspace{\tabcolsep}
\extracolsep{\fill}}lcccc}
  \hline\hline
   Sequence &  \multicolumn{2}{c}{Field center} & Effective exposure
& SIS CCD chip$^\ast$ \\%[-7pt] 
   number & $\alpha$(J2000) & $\delta$(J2000) &
[GIS/SIS] (ks) & \\ \hline 
   5502700 \dotfill & 
$20^{\rm h}03^{\rm m}12^{\rm s}$ &
$32^\circ43^{\prime}56^{\prime\prime}$ & 
7/10 & S0C1/S1C3 \\
   & & & & S0C2/S1C0 \\ 
   5502710 \dotfill & 
$20^{\rm h}00^{\rm m}51^{\rm s}$ &
$32^\circ49^{\prime}25^{\prime\prime}$ & 
8/9 & S0C1/S1C3 \\
   & & & & S0C2/S1C0 \\ 
   5502720 \dotfill & 
$20^{\rm h}01^{\rm m}34^{\rm s}$ &
$32^\circ30^{\prime}55^{\prime\prime}$ & 
11/11 & S0C0/S1C2 \\
   & & & & S0C1/S1C3 \\ \hline
 \end{tabular*}
\vspace{6pt}\par\noindent
 $*$ ``S0C1/S1C3'' is an abbreviation for ``SIS0-chip1/SIS1-chip3''.
\end{table*}

% Fig.1
 \begin{figure*}[t]
  \begin{verse}
   \centerline{\psbox[xsize=1#1]{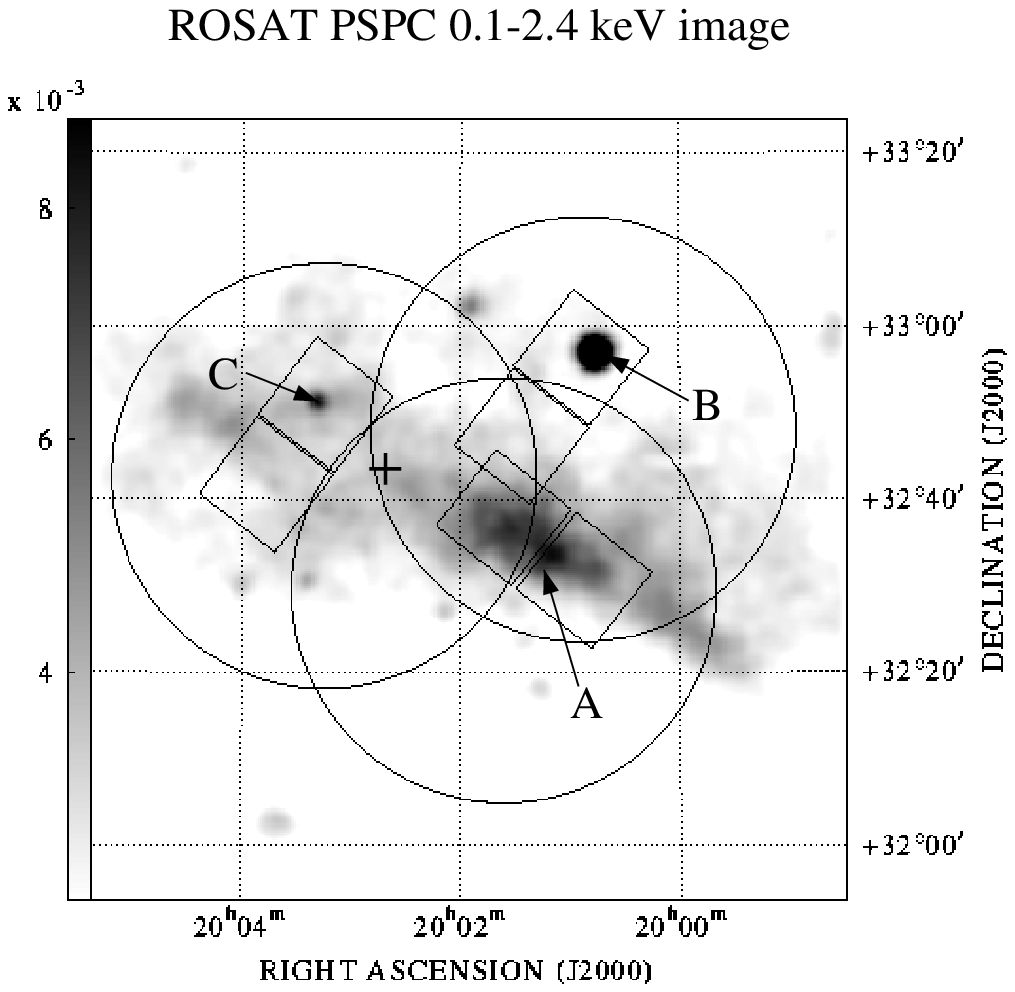}}
   {\footnotesize \setlength{\baselineskip}{9.5pt}%
   Fig.~1.\hspace{4pt}ROSAT PSPC image with the FOVs of the ASCA observation. 
   The gray-scale bar is in units of counts s$^{-1}$ arcmin$^{-2}$.
   The circles and the squares represent the GIS and SIS FOVs,
   respectively. The cross shows the position of G69.7+1.0 obtained
   with the radio observation (Green 1998).

   }
  \end{verse}  
 \end{figure*}

Observations with the ASCA satellite (Tanaka et al.\ 1994) 
in the direction of G69.7+1.0 were performed 
on 1997 November 20--21.  We
selected 3 positions, listed in table~1, 
in order to cover the entire emission based on
the ROSAT observation.
ASCA has four detectors mounted in the 
focal plane of its telescope.
Two of the detectors are
solid-state imaging spectrometers (SIS~0 and SIS~1) (Yamashita et
al. 1997), which were working in the 2-CCD Faint mode.  The other two XRT
are equipped with gas-imaging spectrometers 
(GIS~2 and GIS~3) (Ohashi et al.\ 1996; Makishima et al.\ 1996), 
which were working in the PH mode.  
While the GISs have a field of view (FOV) of
50$^{\prime}$ in diameter, the SISs have a rectangular $11^{\prime} \times
22^{\prime}$ FOV in the 2-CCD mode.  Figure~1 shows
the ASCA FOV superposed on the ROSAT PSPC image.
Detailed information concerning the observation are also listed in table~1.

% Fig.2
  \begin{figure*}[t]
   \begin{verse}
    \centerline{\psbox[xsize=1#1]{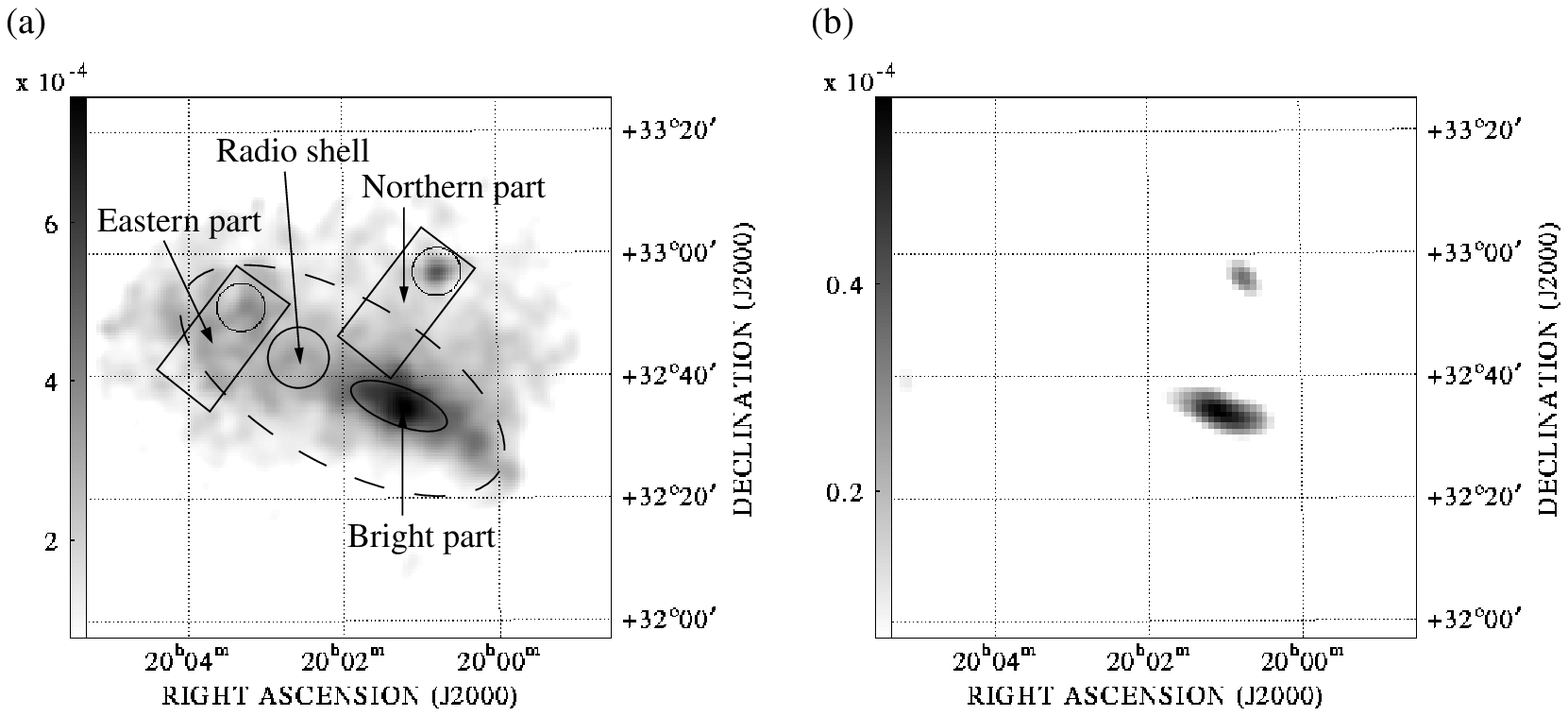}}
    {\footnotesize \setlength{\baselineskip}{9.5pt}%
    Fig.~2.\hspace{4pt}GIS mosaic images 
    observed in the 0.7--1.5 keV band(a) and in the 1.5--5 keV band(b).
    The scale bars are in units of counts s$^{-1}$ arcmin$^{-2}$.
    The rectangles and circles in figure 2a indicate 
    the regions from which the spectra
    displayed in figure~3 were extracted.

    }
   \end{verse}  
  \end{figure*}

We rejected the data at the South Atlantic Anomaly (SAA), 
the high-background regions with the geomagnetic 
cut-off rigidities of $<$ 6 GeV/{\it c}. 
We also excluded the data at low-elevation angles 
from the Earth rim.
The data screening criteria are  
$< 5^{\circ}$ from the night earth rim,
$< 25^{\circ}$ from the bright earth rim for the GIS,
$< 45^{\circ}$ for the SIS~0,
$< 20^{\circ}$ for the SIS~1.
As for the GIS data, we applied a ``flare-cut'' to maximize the 
signal-to-noise ratio described in Y.\ Ishisaki et al.\
(1997 ASCA News 5, 26).
SIS data were acquired in the 2-CCD mode in 1997, and hence
the residual dark distribution (RDD) would become a problem.
Therefore, we applied the RDD correction to the SIS data.
Then, they were corrected for any dark-frame error, the echo
effect and any charge-transfer inefficiency
(T.\ Dotani et al.\ 1995 ASCA News 3, 25, 1997 ASCA News 5, 14).
Hot and flickering pixels were also removed.

The background consists of 
the non-X-ray background and the cosmic X-ray background.
For the GIS, the non-X-ray background was reproduced with a 
method introduced by Ishisaki (1996).
The cosmic X-ray background data were extracted 
from the Large Sky Survey (Ueda et al.\ 1999) data
observed during the ASCA PV phase.
We subtracted these background data from the GIS data.
For the SIS, we subtracted the blank-sky data as background.
We estimated the contribution of the galactic ridge emission 
to be negligible (Yamauchi, Koyama 1993).
In a spectral analysis,
we binned the data into groups having a minimum of 20 counts per bin,
and subtracted the background spectrum,
which was grouped to match the source data.

As for the ROSAT PSPC observation, 
we retrieved the ROSAT archival data through the HEASARC/GSFC
Online Service.
The sequence ID of the data used here is RP500323.
The pointing position was ($\alpha$, $\delta$)(J2000) = 
($20^{\rm h}02^{\rm m}40^{\rm s}\hspace{-4pt}.\hspace{2pt}8$,
$32^\circ43'48.\hspace{-2pt}''0$) and 
the net exposure time was 12 ks.
The spectra were rebinned as well as those of the ASCA data.
The background spectrum was determined from a circular region of 
radius 34$^{\prime}$ at the center of the PSPC FOV,
excluding both the X-ray arc and the point sources.

\section{Results}

\subsection{X-Ray Image}

We generated exposure-corrected, background-subtracted images of the
GIS.  Figure~2 shows GIS mosaic images both 
in the soft band (0.7--1.5 keV) and in the hard band (1.5--5 keV).  
The soft band image (figure~2a) shows an extended X-ray emission 
which is similar to the extended source ``A'' seen
in the ROSAT PSPC image (figure~1).
The other brightest sources in figure~1 
are point sources ``B'' and ``C''.
Source C is missing in the ASCA image.  
Since the PSPC count rate of source C is $\sim 30$\% of that of source B,
the expected fluxes of source C for the GIS and SIS
are considered to be below the threshold. 
The hard band image (figure~2b) shows only two sources: 
the northern source is a point source from the ASCA PSF of the XRT,
whereas the southern source is surely 
an extended source. 

The northern point source in figures 2a and b
was identical to source B seen in the ROSAT PSPC observation.
Source B corresponds to 
the ROSAT All-Sky Survey Bright Source Catalogue
(RASS-BSC) source 
1RXS J200047.0+325708 at position
($\alpha$, $\delta$)(J2000) =
($20^{\rm h}00^{\rm m}47^{\rm s}\hspace{-4pt}.\hspace{2pt}0$, 
$32^\circ57'08.\hspace{-2pt}''5$),
and it is within 20$^{\prime\prime}$ from a K2 star, 
HD~189806 (TYC 2674$-$1400$-$1), to which
the distance is 56 pc (Voges et al.\ 1999).

% Table 2
\begin{table*}[t]
\begin{center}
Table~2.\hspace{4pt}Best-fit parameters of the thin thermal model 
(VMEKAL model).$^\ast$
\end{center}
\vspace{6pt}
 \begin{tabular*}{\textwidth}{@{\hspace{\tabcolsep}
\extracolsep{\fill}}lccccc}
  \hline\hline
  Parameter & Bright part & Eastern part & Northern part & Radio
shell & Entire$^\dagger$\\ 
  \hline
   $N_{\rm H}$ ($10^{21}$ cm$^{-2}$)\dotfill & 1.9$\pm 0.4$ &
0.7$\pm 0.1$ & 2.5$^{+0.7}_{-0.6}$ & 1.0$^{+1.5}_{-0.3}$ & 
1.6$^{+0.5}_{-0.4}$\\
  $kT_{\rm e}$ (keV) \dotfill & 0.45$\pm0.03$ & 0.46$^{+0.03}_{-0.02}$ &
0.40$^{+0.04}_{-0.03}$ & 0.47$^{+0.08}_{-0.13}$ & 
0.42$^{+0.05}_{-0.04} $ \\[4pt]
  \multicolumn{5}{l}{Abundance (relative to cosmic value)} \\
  \hspace*{15pt}Si \dotfill & 0.3$\pm 0.2$ &
  0.9$^{+0.4}_{-0.3}$ & 0.8$\pm 0.4$ & $<$ 2.0 & 
0.6$\pm 0.2$ \\
  \hspace*{15pt}Fe \dotfill & 0.07$\pm 0.03$ &
0.07$\pm 0.03$ & $<$ 0.09 & 0.16$^{+0.09}_{-0.10}$ & 
0.13$\pm 0.04$ \\
  $\chi^{2}$/d.o.f \dotfill & 311.8/269 & 315.7/274 & 288.1/261 & 98.2/132 & 
685.6/445\\
  \hline
 \end{tabular*}
\vspace{6pt}\par\noindent
$\ast$ Errors on parameters are computed using $\Delta\chi^{2}$ = 2.7.
\vspace{6pt}\par\noindent
$\dagger$ Dashed elliptic region shown in figure~2a.
\end{table*}

We extracted the spectra of this source from a 3$^{\prime}$ radius 
circle for the GISs, SISs and ROSAT PSPC.
In order to subtract the component of extended emission from the data,
the background spectra were determined from a 4$^{\prime}$--8$^{\prime}$ radius
annular region for the GISs and PSPC, and from
the FOV of the CCD chip S0C1/S1C3 excluding a 4$^{\prime}$ radius 
circle of the source for the SISs.
Since it has been reported that SIS data 
tend to infer a higher column density 
(ASCA Guest Observer Facility's home page at 
{\it http://heasarc.gsfc.nasa.gov/docs/asca/cal\_probs.html}),
we used SIS data above 0.8~keV.

We performed a combined-fit using five data sets, GIS~2, GIS~3, SIS~0,
SIS~1, and PSPC data. 
Since we could expect coronal X-ray emission from the K2 star, we fitted
them with a thin thermal model (MEKAL model, Mewe et al.\ 1985; Liedahl
et al.\ 1995) of cosmic abundances (Anders, Grevesse 1989).  
We used the model of Morrison and McCammon (1983) 
for the interstellar absorption.
The parameter values that provide a minimum $\chi^{2}$ were taken as 
``best-fit'' parameters, and errors on parameters were computed
using $\Delta\chi^{2}$ = 2.7.
Although the MEKAL model gave
only a poor fit with $\chi^{2}$/d.o.f = 198.8/99,
we obtained the best-fit values for 
the absorption feature, $N_{\rm H}$ $< 3\times10^{18}$ cm$^{-2}$ 
and the electron temperature, $kT_{\rm e}$ = 0.28$\pm0.01$~keV, respectively. 
The obtained value for $kT_{\rm e}$ is typical for 
the temperature of the coronal plasma of late-type stars (Schmitt et al.\ 1990).

\subsection{X-Ray Spectra}

We extracted the spectra from those regions where the SIS observed.
There are three regions: an eastern part, a northern part and 
a bright part.  
As for the data of the eastern part and the northern part,
we excluded the regions 
(marked with circles inside the SIS rectangles in figure~2a)
within $4^{\prime}$ of sources B and C
seen in the ROSAT PSPC image.
In addition to these three regions,
we obtained the spectrum of the region 
which corresponds to the bright region in radio, 
shown in figure~2a as a circle.
The circle has a diameter of $10^{\prime}$;
since it is out of the SIS FOVs, we used the GIS data for this spectrum.
The ASCA and ROSAT PSPC spectra of all regions are shown in figure~3.

We tried to fit spectra with a MEKAL model with variable metallicity,
modified by the interstellar absorption.
In this case, there are residuals at around 0.8--1.3~keV and 1.8~keV.
Therefore, we used a VMEKAL model and allowed the abundances of Si and Fe
to vary during the fitting.
The abundances of other elements were set to be cosmic values.
The best-fit curves are shown by the solid lines in figure~3,
and table~2 summarizes the best-fit parameters.
Although the eastern part has
a slightly lower $N_{\rm H}$ value than other regions,
there is no significant variation in the
electron temperature over our FOV .

% Fig.3
 \begin{figure*}[t]
  \begin{verse}
   \centerline{\psbox[xsize=1#1]{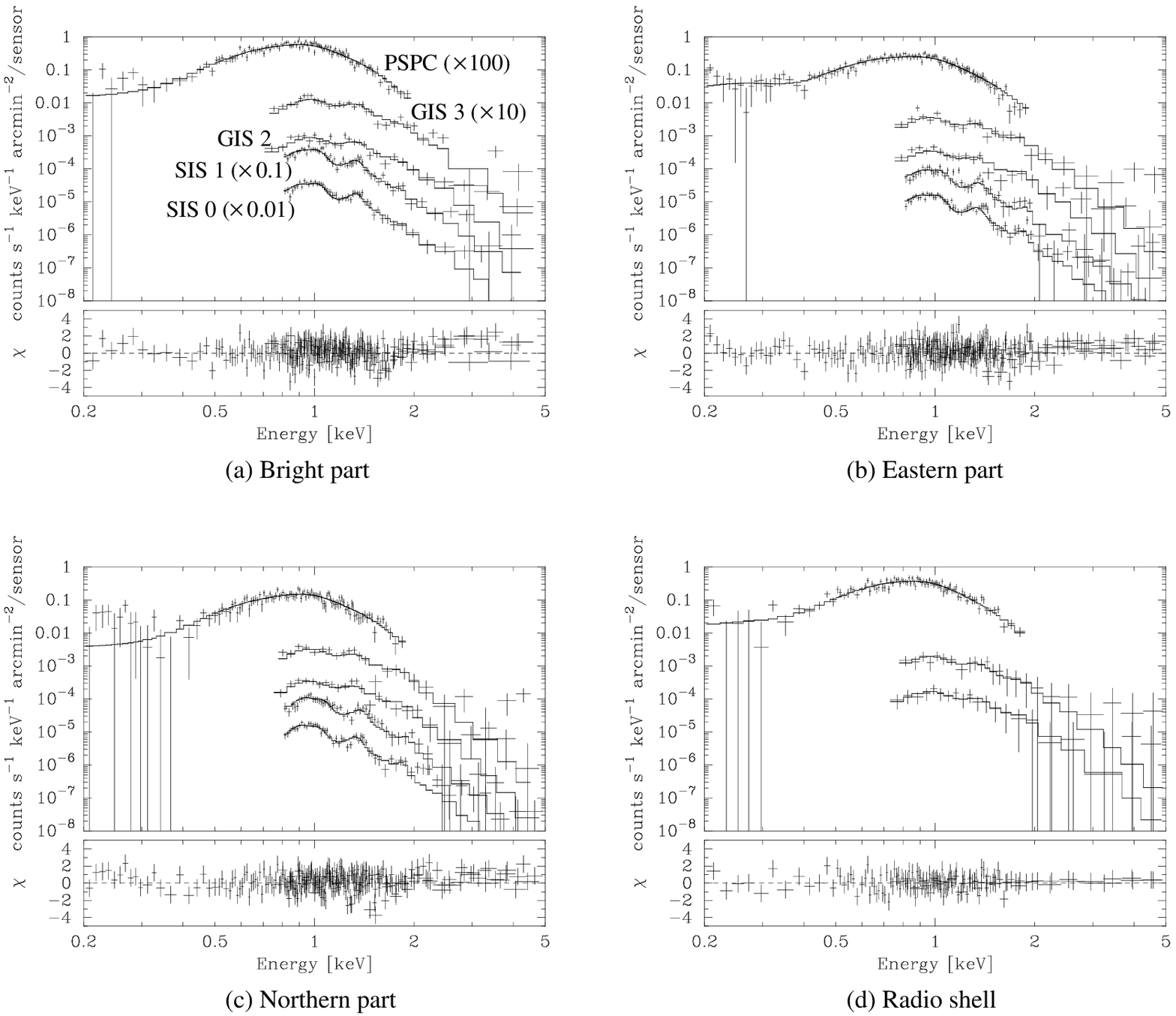}}
   {\footnotesize \setlength{\baselineskip}{9.5pt}%
   Fig.~3.\hspace{4pt}Top panels show the X-ray spectra
   obtained with the ASCA and
   ROSAT PSPC in each case. Best-fit curves are shown with solid lines. 
   (a)--(c) From top to bottom: PSPC ($\times$100), GIS~3 ($\times$10),
   GIS~2, SIS~1 ($\times$0.1), SIS~0 ($\times$0.01),
   (d) PSPC ($\times$100), GIS~3 ($\times$10), GIS~2.
   The lower panels show the residuals of the fits
   for only GIS~2 and SIS~0 besides the PSPC.

   }
  \end{verse}
 \end{figure*}

In figure~3, 
the remaining residuals above 2~keV can be seen 
at the spectrum of the bright part,
which may indicate an additional non-thermal component.
We then fitted the spectrum with a two-component model consisting of
a thin thermal model and a power-law model.
Since we could not determine all of the parameters
when the electron temperature  
and the abundances of Si and Fe are free parameters,
we fixed these parameters to the values which we obtained 
with the fitting using only a thin thermal model.
However, this attempt could not produce a
significant improvement to the fit.

\subsection{Timing Analysis}
We searched for pulsations using GIS data 
within $4^{\prime}$ from the peak of the bright region.
We used both high and medium bit-rate data.
After barycentering the photon times of arrival, 
the power spectral density was calculated by the 
fast Fourier-transformation method.
The Nyquist frequency was set at 8 Hz.
We found no significant pulsation.
Other regions in our FOV do not have sufficient counting statistics to
perform a reliable timing analysis.

\section{Discussion}

We compared the X-ray emission with the radio emission 
in the direction of G69.7+1.0.
Figure~4 shows the ROSAT PSPC image 
superimposed on the 4850 MHz radio contours.
The radio map was obtained through the Skyview
supported by HEASARC/GSFC.
The X-ray emission observed by the PSPC extends beyond 
the radio shell of G69.7+1.0, which is 
centered at ($\alpha$, $\delta$)(J2000) = 
($20^{\rm h}02^{\rm m}42^{\rm s}$,
$32^\circ43^{\prime}28^{\prime\prime}$).
The X-ray peak occurs
roughly at ($\alpha$, $\delta$)(J2000) =
($20^{\rm h}01^{\rm m}10^{\rm s}$,
$32^\circ35^{\prime}$), well outside the radio shell, 
at $25^{\prime}$ from its center.
Usually, the radio shell of an SNR
appears to correspond to, or surround, its 
X-ray bright region (Seward 1990; Green 1998).
Therefore, it is unlikely that the bright part of the X-ray emission 
is associated with G69.7+1.0, but it is still probably an SNR,
because of its thermal emission and extended morphology.
We suggest that the observed X-ray emission may be a new SNR, 
different from the radio SNR, G69.7+1.0,
and have named it AX J2001$+$3235, or G69.4$+$1.2.

% Table 3
\begin{table*}[t]
\begin{center}
Table~3.\hspace{4pt}Parameters for evolved SNRs.\\
\end{center}
\vspace{6pt}
 \begin{tabular*}{\textwidth}{@{\hspace{\tabcolsep}
\extracolsep{\fill}}lccccc}
  \hline\hline
  SNR\hspace*{80pt} & Distance & $kT_{\rm e}$ & Radius
   & $E_{\rm t}$ & Age \\%[-7pt]
   & (kpc) & (keV) & (pc) & ($10^{49}$ erg) & ($10^{4}$ yr)\\ \hline
  AX J2001$+$3235 \dotfill & 2.5 & 0.42$^{+0.05}_{-0.04}$
& 22 & $2.6^{+0.3}_{-0.2}$ & $3.7\pm 0.3$ \\ 
  \hspace*{87pt}\dotfill & 14.4 & 0.42$^{+0.05}_{-0.04}$
& 128 & $(2.1\pm 0.2)\times10^{2}$ & $4.9\pm 0.4$ \\ \hline 
  Cygnus Loop \dotfill & 0.77$^\ast$ & 0.26$^\ast$ 
& 18.8$^\ast$ & $5$ & 2$^\ast$ \\ 
  Vela SNR \dotfill & 0.25$^\dagger$ & 0.11, 0.45$^{\ddagger, \sharp}$ 
& 17$^\ddagger$ & $2$ & 1$^\ddagger$ \\ 
  Puppis A \dotfill & 2.2{\small $^{\S}$} & 0.60{\small $^{\S}$} 
& 8.5{\small $^{\S}$} & $4$ & 0.4{\small $^{\S}$} \\ 
  \hline
 \end{tabular*}
\vspace{6pt}\par\noindent
$\ast$ Ku et al.\ (1984)
\vspace{6pt}\par\noindent
$\dagger$ Cha et al.\ (1999)
\vspace{6pt}\par\noindent
$\ddagger$ Bocchino et al.\ (1999)
\vspace{6pt}\par\noindent
{\small {\S}} Gorenstein et al.\ (1974)
\vspace{6pt}\par\noindent
$\sharp$ Bocchino et al.\ (1999) used a two-temperature emission model.
\end{table*}

% Fig.4
  \begin{figure*}[t]
   \begin{verse}
    \centerline{\psbox[xsize=1#1]{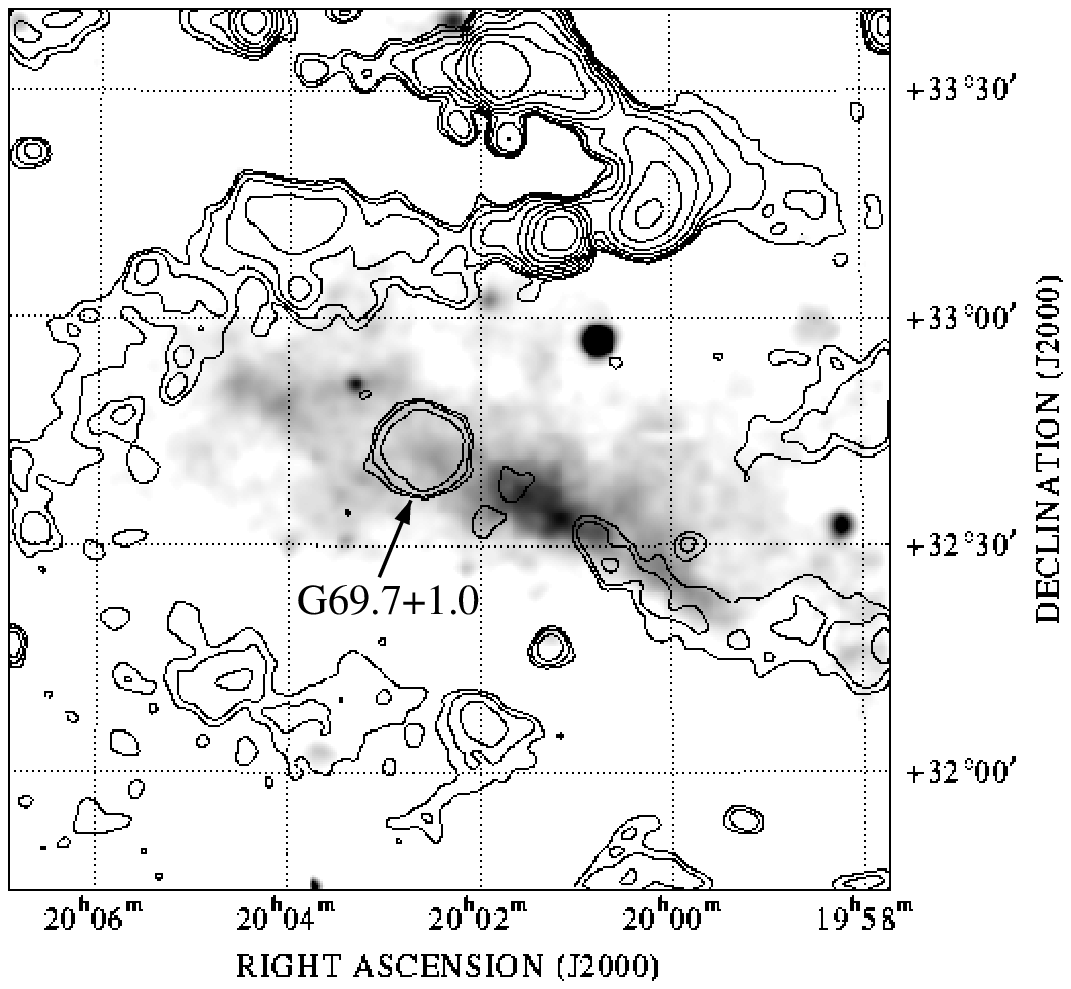}}
    {\footnotesize \setlength{\baselineskip}{9.5pt}%
    Fig.~4.\hspace{4pt}ROSAT PSPC image of the vicinity of G69.7+1.0 
    with an overlay of 4850 MHz survey contour map.
    The contour levels are 5, 10, 20, 40, 80, 160,
    320 and 640 in units of mJy/beam.

    }
   \end{verse} 
  \end{figure*}

% Fig.5
 \begin{figure*}[t]
  \begin{center}
   \psbox[xsize=1#1]{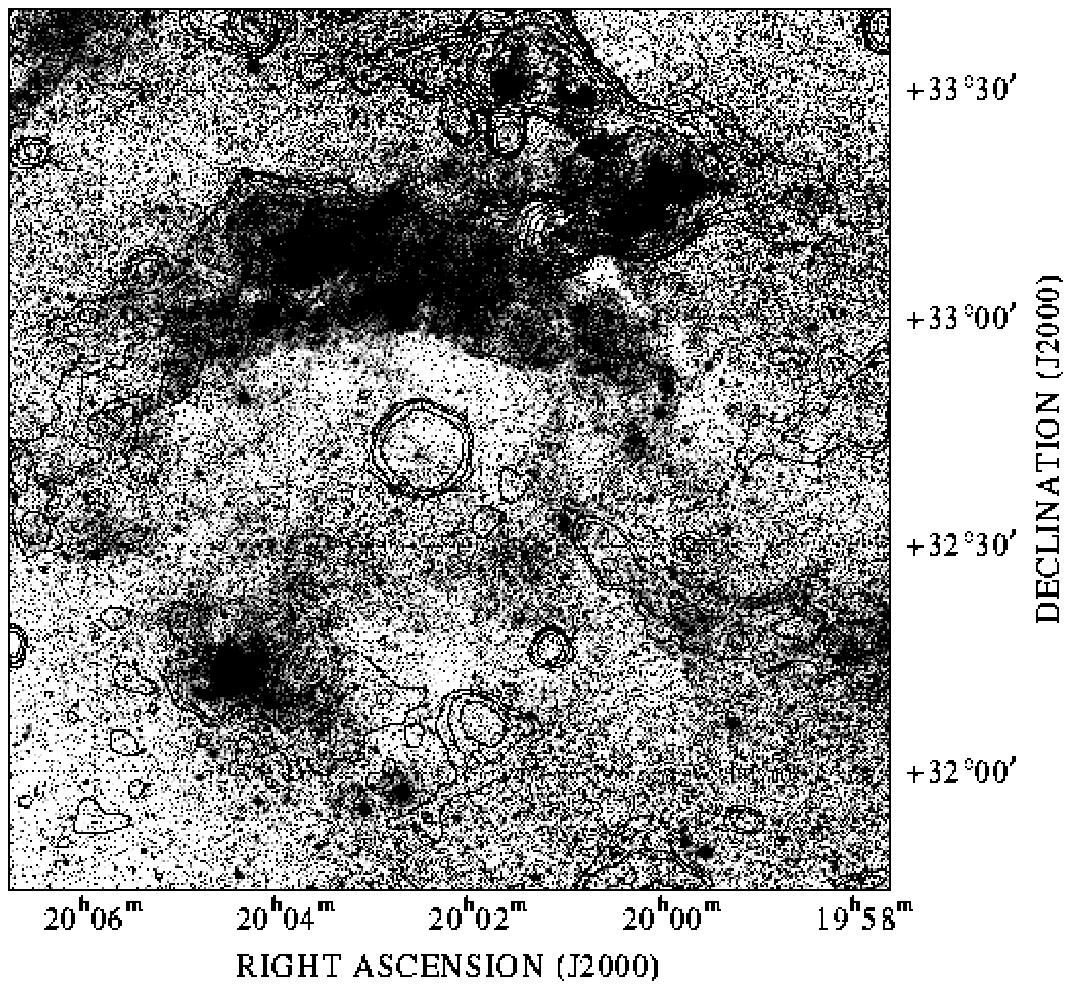}\\
   {\footnotesize Fig.~5.\hspace{4pt}Digitized Sky Survey image of the
   same region as figure 4 with an overlay of 4850 MHz survey contour map.}
  \end{center}  
 \end{figure*}

The radio map shows that a broken shell structure seems to surround
the extended X-ray emission.
This shell has a diameter of about $1^{\circ}$,
as shown in figure~4.
We also found a corresponding optical shell
in the Digitized Sky Survey image 
(retrieved from HEASARC/GSFC, figure~5).
Even though additional observations would be required to establish
whether the optical and radio emission is indeed an SNR,
we note that this structure is remarkably similar to that observed in
known SNR, as for instance, Cygnus Loop (Ku et al.\ 1984).
Therefore, it cannot be excluded that
this structure represents a shell of an evolved SNR.
The X-ray emission with the temperature of about 0.4~keV 
supports that the SNR is in middle-aged.
No association of the X-ray emission with this large shell 
has been reported so far.
Asaoka et al.\ (1996) suggest that the X-ray emission is an SNR, and
is a part of another shell with a radius of about $3^{\circ}$,
which is different from our conclusion.
Our finding is that the $1^{\circ}$ diameter shell
at optical and radio wavelengths may be
related to the extended X-ray emission of AX J2001$+$3235.

We note that the X-ray emission is elongated to 
the south-west direction where the radio and optical shells are broken. 
Both the elongation in X-rays and the breakout at optical and radio 
wavelengths may be due to the low density of the interstellar medium
in the western part.
This fact confirms that the X-ray emission in the direction of
G69.7+1.0 is associated with an evolved SNR 
having both radio and optical shells,
rather than G69.7+1.0.
Follow-up radio and optical observations
are needed to study more in detail 
the relationship between the X-ray emission and
the $1^{\circ}$ diameter shell.

We have estimated the distance of AX J2001$+$3235 using the measured
column density.
We extracted the GIS and PSPC spectra from the region shown 
as a dashed ellipse in figure~2a and
fitted them with the same model as that described in subsection 3.2.
The fitting result is shown in table~2.
Although the poor fit suggests that the model is too simple,
we employed this result to obtain the average SNR properties
as a first-order estimate.
We compared the column density of AX J2001$+$3235 with that of 
CTB 80 (G69.0+2.7) which is the SNR located $1^{\circ}$ from AX J2001$+$3235. 
The distance to CTB 80 has been measured by various methods 
(dispersion measure of the central pulsar, kinematic estimates from
the detected infrared shell, and the optical reddening estimates),
and 2.5~kpc is commonly used (Blair et al.\ 1984; Case, Bhattacharya 1998).
On the other hand, the best-fit value of 
the column density $N_{\rm H}$ of CTB 80 obtained with 
the ROSAT observation was
$3.0\times10^{21}$ cm$^{-2}$ (Safi-Harb et al.\ 1995).
The similar value of the column density of AX J2001$+$3235
$N_{\rm H}$ $\sim2\times10^{21}$ cm$^{-2}$ 
suggests the similar distance of $\sim$ 2.5~kpc.

Employing the distance of 2.5~kpc, 
we calculated the parameters of AX J2001$+$3235.
The fitting result of the spectra of a dashed elliptic region 
gives a lower limit of the thermal energy contained 
within the entire SNR. From the spectral fitting,
we obtained $kT_{\rm e}$ of 0.42 keV and 
EM/(4$\pi d^{2})\sim$ 4.0$\times10^{-16}$ cm$^{-5}$ where EM is 
the emission measure, $\int n_{\rm e}n_{\rm H}\,dV$, 
$n_{\rm e}$ and $n_{\rm H}$ are the mean number density of 
electron and hydrogen, and $d$ is the distance to the SNR.
Assuming that the object is 
an ellipsoid with a major axis of
$0^{\circ}\hspace{-4.5pt}.\hspace{-.5pt}51$ ($\sim$ 22~pc) 
and a minor axis of
$0^{\circ}\hspace{-4.5pt}.\hspace{-.5pt}24$ ($\sim$ 10~pc),
and the number density of hydrogen is
equal to that of electron, 
the derived mean number density of hydrogen 
$n_{\rm H}$ is $\sim$ 0.06 cm$^{-3}$.
The thermal energy $E_{\rm t}$ is calculated to be
\begin{eqnarray*}
E_{\rm t} &\sim& n_{\rm H}VkT_{\rm e} \\
                &\sim& 2.6 \times 10^{49} \left( \frac{d}{2.5~{\rm kpc}} \right)^{2.5} {\rm erg},
\end{eqnarray*}
where $V$ is the volume of the calculated region.

We compared the parameters of AX J2001$+$3235 with those of other evolved SNRs,
the Cygnus Loop (Ku et al.\ 1984), 
Vela and Puppis SNRs (Gorenstein et al.\ 1974; Bocchino et al.\ 1999;
Cha et al.\ 1999).
The results are summarized in table~3.
The age of AX J2001$+$3235 was derived using the Sedov model with an
explosion energy of $10^{51}$~erg.
We can see that all of the parameters 
(electron temperature, radius, thermal energy, age)
of  AX J2001$+$3235 have typical values of evolved SNRs.
We also show the parameters with the assumption of 
the distance of 14.4~kpc, which was obtained 
with the $\Sigma$--{\it D} relation for G69.7+1.0.
In this case, the size and the thermal energy of AX J2001$+$3235
are much larger than other evolved SNRs.
Therefore, a distance as large as 14.4~kpc is very unlikely.

\section{Summary}
We have studied the X-ray emission of a large region
around the known radio SNR G69.7+1.0, and found that
the large extended X-ray source, AX J2001$+$3235, 
is not associated with G69.7+1.0.
This source has an elongated shape with a size of 
$1^{\circ}$, which is much bigger than that of G69.7+1.0. 
It is surrounded by another large shell at both optical and radio wavelengths.
The column density and the electron temperature obtained by the ASCA and
ROSAT PSPC data are 
$N_{\rm H} \sim 2\times10^{21}$ cm$^{-2}$ and $kT_{\rm e}$ $\sim$ 0.4~keV. 
The $N_{\rm H}$ value of AX J2001$+$3235 indicates that it is closer 
than 14.4~kpc, which is the distance to G69.7+1.0 derived 
from the $\Sigma$--{\it D} relation.
We estimate the distance to AX J2001$+$3235 to be $\sim$ 2.5~kpc.
The radio and optical shells and
the derived parameters of AX J2001$+$3235 
using the fitting results of the X-ray spectra
are consistent with the idea 
that AX J2001$+$3235 is an evolved, previously unknown, SNR.
\par
\vspace{1pc}\par
The authors are grateful to all the members of the ASCA team. 
We are also grateful to the referee, F.\ Bocchino, for his useful comments.
K.Y.\ is supported by JSPS Research Fellowship for Young Scientists.

\section{References}
\re
Anders E., Grevesse N.\ 1989, Geochim.\ Cosmochim.\ Acta 53, 197

\re
Asaoka I., Egger R., Aschenbach B.\ 1996, MPE Report 263, 233

\re
Blair W.P., Kirshner R.P., Fesen R.A., Gull T.R.\ 1984, ApJ 282, 161

\re
Bocchino F., Maggio A., Sciortino S.\ 1999, A\&A 342, 839

\re
Case G.L., Bhattacharya D.\ 1998, ApJ 504, 761 

\re
Cha A.N., Sembach K.R., Danks A.C.\ 1999, ApJ 515, L25 

\re
Fabian A.C., Willingale R., Pye J.P., Murray S.S.,
Fabbiano G.\ 1980, MNRAS 193, 175 

\re
Gorenstein P., Harnden F.R.\ Jr, Tucker W.H.\ 1974, 
ApJ 192, 661

\re
Green D.A.\ 1998, A Catalogue of Galactic
Supernova Remnants  (1998 September version) (Mullard Radio Astronomy
Observatory, Cambridge)

\re
Ishisaki Y.\ 1996, Ph.D.\ thesis, The University of Tokyo

\re
Ku W.H.-M., Kahn S.M., Pisarski R., Long K.S.\ 1984, ApJ 278, 615

\re
Liedahl D.A., Osterheld A.L., Goldstein W.H.\ 1995, ApJ
438, L115

\re
Makishima K., Tashiro M., Ebisawa K., Ezawa H., Fukazawa Y., Gunji S., 
Hirayama M., Idesawa E.\ et al.\ 1996, PASJ 48, 171 

\re
Mewe R., Gronenschild E.H.B.M., 
van den Oord G.H.J.\ 1985, A\&AS 62, 197 

\re
Morrison R., McCammon D.\ 1983, ApJ 270, 119

\re
Ohashi T., Ebisawa K., Fukazawa Y., Hiyoshi K., Horii M., Ikebe Y.,
Ikeda H., Inoue H.\ et al.\ 1996, PASJ 48, 157 

\re
Reich W., F\"{u}rst E., Reich P., Junkes N.\
1988, in Proc.\ 101st colloquium
Supernova Remnants and Interstellar Medium, 
ed R.S.\ Roger, T.L.\ Landecker
(Cambridge University Press, Cambridge) p293

\re
Reid P.B., Becker R.H., Long K.S.\ 1982, ApJ 261, 485

\re
Safi-Harb S., \"Ogelman H., Finley J.P.\ 1995, ApJ 439, 722

\re
Schmitt J.H.M.M., Collura A., Sciortino S., 
Vaiana G.S., Harnden F.R.\ Jr, Rosner R.\ 1990, ApJ 365, 704

\re
Seward F.D.\ 1990, ApJS 73, 781

\re
Tanaka Y., Inoue H., Holt S.S.\ 1994, PASJ 46, L37 

\re
Ueda Y., Takahashi T., Inoue H., Tsuru T., Sakano M., Ishisaki Y.,
Ogasaka Y., Makishima K.\ et al.\ 1999, ApJ 518, 656 

\re
Voges W., Aschenbach B., Boller Th., Br\"auninger H.,
Briel U., Burkert W., Dennerl K., Englhauser J.\ et al.\ 1999, A\&A 349, 389

\re
Yamashita A., Dotani T., Bautz M., Crew G., 
Ezuka H., Gendreau K., Kotani T., Mitsuda K.\ et al.\ 1997, IEEE Trans.\ Nucl.\ Sci.\ 44, 847 

\re
Yamauchi S., Koyama K.\ 1993, ApJ 404, 620 

\end{document}